\documentclass[epj,referee]{svjour}
\usepackage{graphics}
\begin{document}
\title{A Self--organized model for network evolution}
\subtitle{Coupling network evolution and extremal dynamics}
\author{Guido Caldarelli\inst{1,2,3} \and Andrea Capocci\inst{4}
\and Diego Garlaschelli\inst{5}
}                     
%
%
\institute{INFM-CNR Centro SMC Dipartimento di Fisica Universit\`a 
``Sapienza'' Piazzale Aldo Moro 2, 00185 Roma, Italy.  
\and 
Centro Studi e Ricerche e Museo della Fisica ``Enrico Fermi'' 
Compendio Viminale, 00185 Roma Italy.
\and
Linkalab, Center for Complex Networks Research, Sardegna (Italy).
\and
Dipartimento di Informatica e Sistemistica Universit\`a Sapienza 
via Salaria 113, 00185 Roma, Italy.
\and 
Dipartimento di Fisica, Universit\`a di Siena, Via Roma 56, 53100 Siena, Italy 
}
\date{Received: date / Revised version: date}
%
\abstract{
Here we provide a detailed analysis, along with some extensions and additonal investigations, of a recently proposed  \cite{naturelett} self--organised model for the evolution of complex networks. 
Vertices of the network are characterised by a fitness variable evolving 
through an extremal dynamics process, as in the Bak--Sneppen  \cite{BS} model representing a prototype of Self--Organized Criticality. The network topology is in turn 
shaped by the fitness variable itself, as in the fitness network model \cite{fitness}. 
The system self--organizes to a nontrivial state, characterized by a power--law decay of dynamical and topological 
quantities above a critical threshold. The interplay between topology and dynamics in the system is the key ingredient leading to an unexpected behaviour of these quantities.
} 
\maketitle
\section{Introduction}
\label{intro}
Complex Networks represent a very active topic in the field of Statistical Physics. 
The possibility to describe several different systems as 
structures made of vertices (subunits) connected by edges (their interactions) has proved successful in a variety 
of disciplines, ranging from computer science to 
biology \cite{AB01,book1}. 
One common feature amongst all these different physical systems is 
the lack of a characteristic scale for the degree (the number of 
edges per vertex), and the presence of pairwise correlations between the degrees of neighboring vertices.
On top of these topological properties, also the dynamical processes acting on a network often show unexpected features.
Indeed, the dynamics of the outbreak of an infectious disease 
is totally different when defined on complex 
networks rather than on regular 
lattices \cite{ale,book2}. 

Many different models have been proposed in order to reproduce such 
behaviour. One of the earliest ones, introduced by Barab\'asi and Albert \cite{BA99}, uses ``growth'' and ``preferential attachment'' 
as active ingredients for the onset of degree scale invariance.  
On the other hand, there is growing empirical 
evidence \cite{shares,mywtw} that networks are often shaped 
by some variable associated to each vertex, 
an aspect captured by the `fitness' model \cite{fitness,goh,soderberg}.
In general the onset of scale invariance seems to be related to 
the fine tuning of some parameter(s) in all these models, while it would be 
desirable to present a mechanism through which this feature can develop in a self--organised way.

Here we describe in detail, and partly extend, a recently proposed model  \cite{naturelett} where the topological properties explicitly depend on vertex--specific fitness variables, but at the same time the latter spontaneously evolve to self--organized values, without external fine tuning. 
The steady state 
of the model can be analytically solved for any choice of the connection probability.
For particular and very reasonable choices of the latter, the network develops 
a scale--invariant degree distribution irrespective of the initial condition.

This model integrates and overcomes two separate frameworks that have been explored so far: dynamics on fixed networks and network formation driven by fixed variables. 
These standard approaches to network modelling assume that
the topology develops much faster than other vertex--specific properties 
(no dynamics of the latter is considered), or alternatively that the vertices' properties 
undergo a dynamical process much faster than network evolution 
(the topology is kept fixed while studying the dynamics). 
Of course, by separating the dynamical and topological time--scales, 
and allowing the ``fast variables'' to evolve while the ``slow variables'' 
are fixed (quenched), the whole picture becomes simpler and more tractable. 
However, this unavoidably implies that the slow variables must be 
treated as free parameters to be arbitrarily specified and, more 
importantly, that the feedback effects between topology and dynamics 
are neglected.

The main results highlighted by the self--organised model is that these feedback effects are 
not just a slight modification of the decoupled scenario \cite{naturelett}. 
Rather, they play a major role by driving the system 
to a non equilibrium stationary state with non-trivial properties, a
feature shared with other models of dynamics--topology coupling on 
networks \cite{ref1,ref2,ref3,ginestra,holyst}.

\section{The model}
\subsection{The Bak--Sneppen model and SOC}
\label{sec:1}
In order to define a self--organized mechanism for the onset of scale--invariance, 
we took inspiration from the activity in the field of Self--Organized 
Criticality(SOC) \cite{BTW}. 
In particular, we considered a Bak--Sneppen like evolution for the fitness variable 
\cite{BS}. 
In the original formulation, one deals with a set of vertices (representing biological species) placed on a one--dimensional lattice whose edges represent predation relationships. 
This system is therefore a model for a food chain, and it has been investigated 
to show how the internal dynamics is characterised by scale--free avalanches of evolutionary events \cite{BS}.
Every species $i$ is characterised by a fitness value $x_i$ drawn from a uniform probability distribution 
$\rho(x)$ defined between $0$ and $1$.
The species with the minimum value of the fitness is selected and removed from the 
(eco)system. This reflects the interpretation of the fitness as a measure of success of a particuar species against other species.
Removal of one species is supposed to affect also the others. This is modeled by 
removing the predator and the prey (the nearest neighbours in a 1-dimensional 
lattice) of the species with the minimum fitness.
Three new species are then introduced to replace the three ones that have been removed. 
Alternatively, this event is interpreted as a mutation of the three species towards an evolved state.
The fitnesses of the new species are always extracted from a uniform distribution on the unit interval.
The steady state obtained by iterating this procedure is characterised by a uniform distribution 
of fitnesses between a lower critical threshold $\tau=0.66702 \pm 0.00008$  \cite{Grassberger} and 1. 
The size $s$ of an avalanche, defined as a series of causally connected 
extinctions, follows a scale--free distribution 
$P(s) \propto s^{-\chi}$ where $\chi=1.073 \pm 0.003$  \cite{Grassberger}. 

The behaviour of this model has also been studied on different fixed topologies, including
regular lattices  \cite{BS,gabrie,BSd}, 
random graphs  \cite{BSrn}, small--world  \cite{BSsw} and 
scale--free  \cite{BSsf} networks. The value of the critical threshold $\tau$ is found to depend on the topology, but the stationary fitness distribution always displays the same qualitative step--like behaviour.
These more complicated structures are closer to realistic food webs  \cite{nature}, and allow the number of updated vertices (which equals one plus the degree of the mimum--fitness vertex) to be heterogeneously distributed.
Nevertheless, as long as the network is fixed, the model leads to the ecological paradox that, after a mutation, the new species always inherits exactly all the links of the previous species. This represents a problem, since it is precisely 
the structure of ecological connections among species which is believed to be both the origin and the outcome of macroevolution  \cite{webworld}. At odds with the model, a mutated species is expected to develop different interactions with the rest of the system.

\subsection{Self--organised graph formation}
These problems are overcome if the network is not fixed, and evolves as soon as vertices mutate. At the same time, if the evolving topology is assumed to depend on the fitness values, one obtains a self--organised model that also allows to explore the interplay between dynamical and topological properties.

\begin{figure}
\centerline{
\resizebox{0.9\columnwidth}{!}{%
  \includegraphics{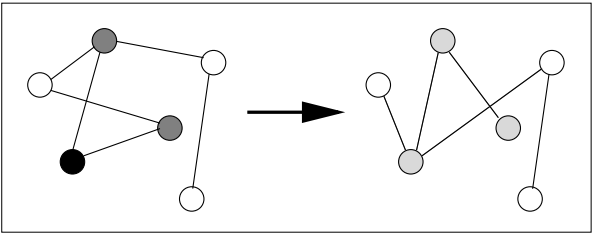}
}
}
\caption{Left: a graph at time $t$. The black vertex has the 
minimum fitness, and the two grey ones are its neighbours. Right: at the next time step $t+1$, three new fitness values are assigned to the vertices (light grey), and connections are established anew between the three vertices and the rest of the network.}
\label{fig:1}       
\end{figure}

The simplest way to put these ingredients together is to couple the Bak--Sneppen model  \cite{BS} with the fitness network model  \cite{fitness}. 
In the model so obtained  \cite{naturelett}, we start by specifying a fixed number $N$ of vertices, and by assigning each vertex $i$ a fitness value $x_i$ initially drawn 
from a uniform probability distribution between $0$ and $1$. 
This coincides with the Bak--Sneppen initial state.
However, {\em we do not fix the topology of the network}, since it is determined by the fitness values themselves as the system evolves.
More specifically, the edge between any pair of vertices $i$ and $j$ is drawn with a fitness dependent probability $f(x_i,x_j)$ as in the fitness model  \cite{fitness}.
At each timestep the species with lowest fitness and all its neighbours 
undergo a mutation, and their fitness values are drawn anew. 
As soon as a vertex $i$ mutates, the connections between it and all the other vertices $j$ are drawn anew with the specified probability $f(x_i,x_j)$.
An example of this evolution for a simple network is shown in Fig.~\ref{fig:1}.

\subsection{Two rules for mutations}
There are various possible choices for updating the fitness of a mutating vertex. In ref. \cite{naturelett} we assumed that each neighbour $j$ of the minimum--fitness vertex receives a completely new fitness drawn  
from the uniform distribution on the unit interval. We shall refer to this choice as rule 1:
\begin{equation}
x_j(t+1)=\eta
\end{equation}
where $\eta$ is uniformly distributed between $0$ and $1$. In this case, irrespective of the number of other vertices $j$ is connected to, $x_j$ will be completely updated.

Besides this rule, here we consider a weaker requirement in which the fitness of each neighbour $j$ is changed only by an amount proportional to $1/k_j$, where $k_j$ is $j$'s degree. We denote this choice as rule 2:
\begin{equation}
x_j(t+1)=\frac {1}{k_j}\eta + \frac{k_j-1}{k_j}x_j(t)
\label{eq:cicco}
\end{equation}
where again $\eta$ is a random number uniformly distributed between $0$ and $1$. 
Rule 2 corresponds to assume that the fitness $x_j$ is completely modified only if $j$ is connected only to the minimum--fitness vertex (that is, $k_j=1$). If $j$ has $k_j-1$ additional neighbours, a share $(k_j-1)/k_j$ of $x_j$ is unchanged, and the remaining fraction $x_j/k_j$ is updated to $\eta/k_j$. This makes hubs affected less than small--degree vertices. Clearly, it also implies that the probability of connection to all other vertices varies by a smaller amount.

\subsection{Arbitrary timescales for link updates}
As we now show, the stationary state of the model does not change even if arbitrary link updating timescales are introduced. In other words, the results are unchanged even if we allow each pair of vertices to be updated at any additional sequence of timesteps, besides the natural updates occurring when any of the two vertices mutates. Two distinct pairs of vertices may also be updated at different sequences of timesteps. In general, if $t_{ij}$ denotes the time when the pair of vertices $i, j$ is chosen for update, we allow a link to be drawn anew between $i$ and $j$ with probability $f[x_i(t_{ij}), x_j(t_{ij})]$. 
To see that this does not change the model results, we note that the fitness of each vertex $i$ remains unchanged until it is selected for mutation, and this occurs only if $i$ happens to be either the minimum--fitness vertex or one of its neighbours. 
Therefore, for any pair of vertices $i,j$, $x_i(t_{ij})=x_i(t'_{ij})$ and $x_j(t_{ij})=x_j(t'_{ij})$, where $t'_{ij}<t_{ij}$ denotes the time of the most recent (before $t_{ij}$) mutation of either $i$ or $j$. Now, since at this latest update the links between the mutating vertex and all other vertices were drawn anew, it follows that $t'_{ij}$ coincides with the timestep when a connection was last attempted between $i$ and $j$, with probability $f[x_i(t'_{ij}), x_j(t'_{ij})]$. Therefore, between $t'_{ij}$ and $t_{ij}$, a link exists between $i$ and $j$ with probability $f[x_i(t'_{ij}), x_j(t'_{ij})]$. On the other hand, if at time $t_{ij}$ the update is performed, a link will be drawn anew with probability $f[x_i(t_{ij}), x_j(t_{ij})]=f[x_i(t'_{ij}), x_j(t'_{ij})]$. Therefore the connection probability is unchanged by the updating event. Since the stationary state of the model only depends on this probability, we find that updating events do not affect the stationary state. 
Therefore our model is very general in this respect, and allows 
for rearrangements of ecological interactions on shorter 
timescales than those generated by mutations. 
This extremely important result also means that, if the whole topology is drawn anew at each timestep, the results will be unchanged. This is a very useful property that can be exploited to perform fast numerical simulations of the model.

\section{Numerical results for rules 1 and 2}
In this section we present numerical results obtained by simulating the model using particular choices of the mutation rule and of the connection probability. For rule 1, these numerical results will be confirmed by the analytical results that we present later on.

Traditionally, one of the most studied properties of the BS model on regular lattices 
is the statistics of avalanches characterizing the SOC behaviour  \cite{BS}. 
Nevertheless, as shown in ref. \cite{notsoc}, in presence of long--range  \cite{BSrn} connections as those displayed by our model, the SOC state can be wrongly assessed in terms of the avalanche statistics. Indeed, the absence of spatial correlations questions criticality even in case the avalanches are power--law 
distributed. We then move to the study of other properties of the system 
by considering the fitness distribution $\rho(x)$ 
and the degree distribution $P(k)$ at the stationary state. 

\subsection{Choice of the connection probability\label{choice}}
To to so, we first need to choose a functional form for the connection probability $f(x,y)$. 
The constant choice $f(x,y)=p$ is the trivial case corresponding to a random graph. This choice is asymptotically equivalent to the random 
neighbour variant  \cite{BSrn} of the BS model, the average degree 
of each vertex being $d=p(N-1)\approx pN$ (we drop terms of 
order $1/N$ from now on). This choice is therefore too simple to introduce novel effects. Indeed, the independence of the topology on the fitness introduces no feedback between structure and dynamics.

We therefore make the simplest nontrivial choice. If we require that the fitness--dependent network has no degree correlations other that those introduced by the local properties alone, the simplest choice for $f(x,y)$ is  \cite{newman_origin,likelihood}
\begin{equation}
f(x,y)=\frac{zxy}{1+zxy}
\label{fermi}
\end{equation}
where $z$ is a positive parameter controlling the number of links. Apart for the structural correlations induced by the 
degree sequence  \cite{newman_origin,likelihood}, higher--order properties are 
completely random, as in the \emph{configuration model}  \cite{siam,maslov}. 
Coming back to the ecological meaning of 
the Bak--Sneppen model, the above choice is consistent with the interpretation that the more a species is connected to other species, the more it is 
fit to the whole environment: the larger $x$ and $y$, the larger $f(x,y)$. 
When $z<<1$, the above connection probability reduces to the bilinear choice
\begin{equation}
f(x,y)=zxy
\label{bilinear}
\end{equation}
In this case, a sparse graph is obtained where structural correlations disappear.

\subsection{Stationary fitness distribution}
\begin{figure}
\centerline{
\resizebox{0.9\columnwidth}{!}{%
  \includegraphics{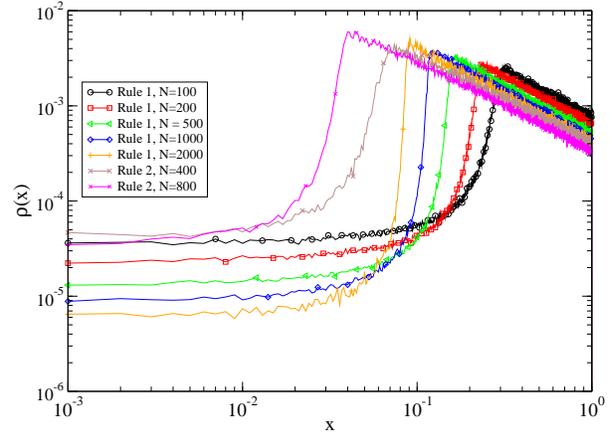}
}
}
\caption{Stationary fitness distribution obtained by numerical simulations of networks with different numbers $N$ of vertices and using either rule 1 or rule 2.}
\label{fig:1bis}       
\end{figure}
In Fig.\ref{fig:1bis} we show the numerical results for a series of computer 
simulations of the model, with sizes ranging between $N=100$ to $N=2000$. Both rule 1 and 2 were used to update the neighbours of the minimum--fitness vertex.
Some qualitative features can be spotted immediately. Firstly, all fitness values move above a positive threshold $\tau$, as in the traditional Bak--Sneppen model on various topologies.
However, here the fitness distribution above the threshold is not uniform. The fitness values self--organize to a scale--free
distribution with the same exponent $-1$ regardless of the size $N$ of the system. This is a remarkable difference, and highlights the effect of the interplay between dynamics and topology.
Finally, rule 1 and 2 produce a similar fitness distribution. 
In section \ref{analytical}, where we provide an analytical solution of the model for rule 1, we demonstrate that the exponent is exactly $-1$.

\subsection{Stationary degree distribution\label{degree}}
\begin{figure}
\centerline{
\resizebox{0.9\columnwidth}{!}{%
  \includegraphics{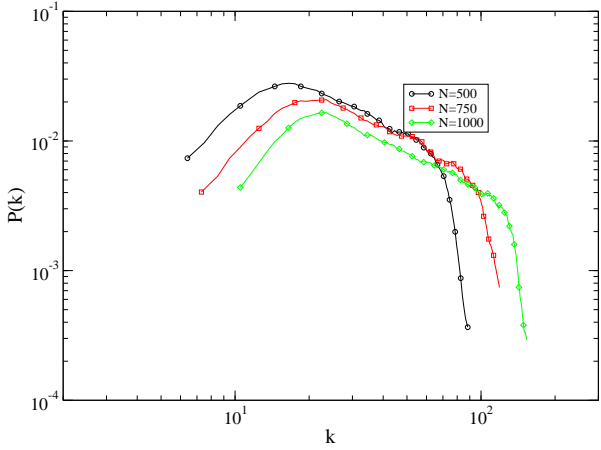}
}
}
\caption{Stationary degree distribution obtained by numerical simulations of networks with different numbers $N$ of vertices.}
\label{fig:1tris}       
\end{figure}
In Fig. \ref{fig:1tris} we also plot the degree distribution $P(k)$, whose behaviour is similar to that of $\rho(x)$.
In particular, we find that the degree distribution has a power--law shape with the same exponent of the fitness distribution, plus two cut--offs at small and large degrees. 
This is not surprising since one can derive analytically the connection existing between degree 
distribution and fitness distribution in the fitness model \cite{pastor,servedio}. The expected degree $k(x)$ of a vertex depends on its fitness $x$. Thus here the lower cut--off corresponds to the value $k(\tau)$ and is the effect of the fitness threshold. The upper cut--off corresponds instead to the maximum possible value $k(1)$, that depends on the parameter choice.
This behaviour will be confirmed by the theoretical results presented in section \ref{analytical}.

\subsection{Size dependence of the threshold}
\begin{figure}
\centerline{
\resizebox{0.9\columnwidth}{!}{%
  \includegraphics{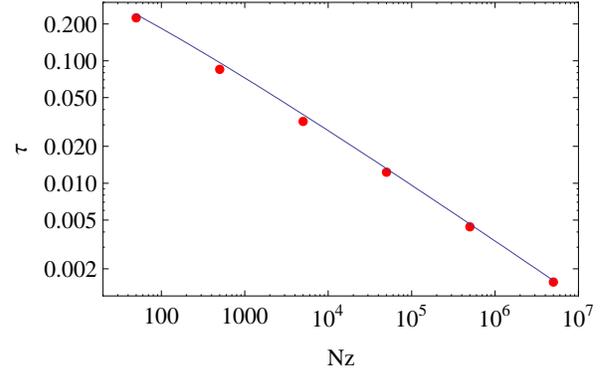}
}
}
\caption{Dependence of the critical threshold $\tau$ on $zN$. The points correspond to numerical simulations (using rule 1) for $N=5000$ and (from left to right) $z=0.01,0.1,1,10,100,1000$. The solid curve is the theoretical prediction $\tau(zN)$ derived in eq.(\ref{eq_tau}) using the analytical solution presented in section \ref{analytical}.}
\label{fig_tau}
\end{figure}

It is clear from Fig.\ref{fig:1bis} that the different sizes affect the value of the threshold. More precisely, the latter decreases as system size increases.
Furthermore, $\tau$ also depends on the parameter $z$.
To better characterise this behaviour, we plot in Fig.\ref{fig_tau} the dependence of $\tau$ on $zN$. We find that $\tau$ only depends on the combination $zN$, a result that we later confirm analytically. Indeed, the solid curve superimposed to the data is obtained using the analytical results of section \ref{analytical}. 

\subsection{Average fitness versus threshold}
\begin{figure}
\centerline{
\resizebox{0.9\columnwidth}{!}{%
  \includegraphics{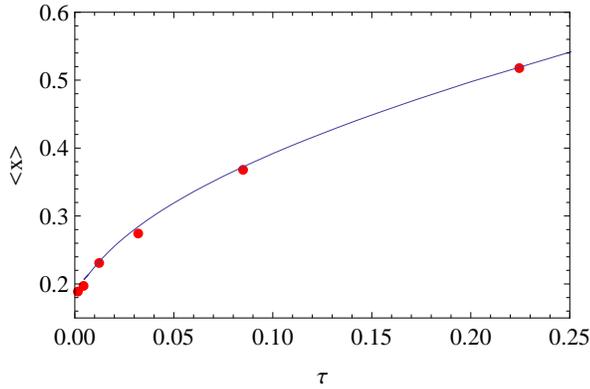}
}
}
\caption{Dependence of the average fitness $\langle x\rangle$ of vertices at the stationary state on $\tau$. The points correspond to numerical simulations (using rule 1) for $N=5000$ and (from right to left) $z=0.01,0.1,1,10,100,1000$. The solid curve is the theoretical prediction derived using the analytical solution presented in section \ref{analytical}.}
\label{fig_mean}
\end{figure}

An overall measure of the evolution of the fitness values from the initial state to the stationary one can be obtained in terms of the average fitness $\langle x\rangle$ of vertices at the stationary state. 
In the initial state, clearly $\langle x\rangle=1/2$. As the fitness values move above the emerging threshold, $\langle x\rangle$ increases, until in the stationary state it reaches an asymptotic value. 
An important trend we identify is that, for fixed $N$, as $z$ increases $\langle x\rangle$  decreases. Since also $\tau$ decreases as $z$ increases, it is interesting to monitor this effect by plotting $\langle x\rangle$ as a function of the threshold $\tau$. This is reported in fig.\ref{fig_mean} for $z=0.01,0.1,1,10,100,1000$. Once again, the simulations agree with the theoretical predictions that will be derived in the next section.

\section{Theoretical results for rule 1\label{analytical}}
If rule 1 is adopted, it is possible to derive complete analytical results even for an arbitrary linking function \cite{naturelett}. 
We first briefly report the general analytical solution, and then show its particular form for specified choices of $f(x,y)$.

\subsection{General analytic solution}
The method we used is to consider the  
master equation for the fitness distribution $\rho(x,t)$ in the system at 
the stationary state \cite{naturelett}. 
If a stationary state exists then
\begin{equation}
 \frac{d\rho(x,t)}{dt}=0 \mbox{      for } t \rightarrow \infty
\end{equation}
This equation can be written in the same limit as 
\begin{equation}
\frac{d\rho(x,t)}{dt}=r^{in}(x,t) -r^{out}(x,t)=0
\end{equation}
where $r^{in,out}(x,t)$ denotes the fraction of vertices whose fitness is $x$ 
entering/leaving the system respectively at time step $t$.
At the steady state ($t\rightarrow \infty$) these quantities no longer depend on  $t$, and we denote them by 
$r^{in}(x)$ and $r^{out}(x)$.
It can be shown \cite{naturelett} that these quantities can be computed separately in terms of the distribution $q(m)$ of the minimum fitness. One finds
\begin{equation}
r^{in}(x) = \frac{1 +\langle k_{min}\rangle} {N} 
\end{equation}
where $\langle k_{min}\rangle\equiv \int_0^\tau q(m) k(m)dm$ is the expected degree of the vertex with minimum fitness. This relation simply states that when the minimum is selected, then one vertex (the minimum itself) plus its neighbours (on average $\langle k_{min}\rangle$) are replaced by new vertices with fitnesses extracted from a uniform distribution. 

One can also show that
\begin{equation}
r^{out}(x) = \left\{
\begin{array}{ccc}
q(x)/N & & x<\tau \\
\rho(x)\int_0^\tau q(m) f(x,m)dm & & x>\tau \\
\end{array}
\right .
\end{equation}
where $\tau$ is defined here as the fitness value below which, in the large size limit, the fitness distribution $\rho(x)=q(x)/N$ is essentially determined by the distribution of the minimum, and above which $\rho(x)>q(x)/N$, or in other words
\begin{equation}
\lim_{N\to\infty}\frac{N\rho(x)}{q(x)}\left\{\begin{array}{ll}=1&x\le\tau\\
>1&x>\tau\end{array}\right.
\label{eq_limit}
\end{equation}

If one now requires $r^{in}(x)=r^{out}(x)$ at the stationary state, it is straightforward to obtain the analytical solution for any form of $f(x,y)$ \cite{naturelett}:
\begin{equation}
\rho(x)=\left\{\begin{array}{ll}
(\tau N)^{-1} &x<\tau\\
\displaystyle{\frac{1}{N\int_0^\tau f(x,m)dm}}
&x>\tau
\end{array}\right.
\label{eq_rho2}
\end{equation}
As a remarkable result, here we find that $\rho(x)$ is in general not uniform for $x>\tau$, in contrast with the Bak--Sneppen model on fitness--independent networks.

The value of $\tau$ is determined implicitly through the normalization condition $\int_0^1\rho(x)dx=1$, which implies
\begin{equation}
\int_\tau^1\displaystyle{\frac{dx}{\int_0^\tau f(x,m)dm}}=N-1
\label{eq_norm}
\end{equation}

The topology of the network at the stationary state can be completely determined 
in terms of $\rho(x)$ as in the static fitness model  \cite{fitness,pastor,servedio}. For instance, the expected degree of a vertex with fitness $x$ is $k(x)=N\int f(x,y)\rho(y)dy$. 
Therefore the above solution fully characterizes both the dynamics and the topology at the stationary state.

As in the standard BS model, 
in the infinite size limit $N\to\infty$ the distribution 
$q(m)$ of the minimum fitness is uniform between $0$ and $\tau$, 
while almost all other values are above $\tau$ \cite{naturelett}. In other words, $q(m)=\Theta(\tau-m)/\tau$.

\subsection{Fitness--independent networks: random graphs}
The latter result generalizes what is obtained for the random--neighbour variant of the BS model \cite{BSrn}.
Indeed, as already noticed, the random--neighbour model is a particular case of our model, obtained with the trivial choice $f(x,y)=p$. The network is a random graph independent of the fitness values, with Poisson--distributed degrees. We now briefly discuss this case as a null reference for more complicated choices discussed below.
Our analytical result in eq.(\ref{eq_rho2}) becomes
\begin{equation}
\rho(x)=\left\{\begin{array}{ll}
(\tau N)^{-1} &x<\tau\\
(p\tau N)^{-1}
&x>\tau
\end{array}\right.
\label{eq_rhop}
\end{equation}
We therefore recover the standard result that, at the stationary state, the fitness distribution is step--like, with on average one vertex below the threshold and all other values uniformly distributed between $\tau$ and $1$. 
We note from eq.(\ref{eq_rho2}) that the uniform character above $\tau$ is only possible when $f(x,y)$ is independent of $x$ and $y$, or in other words when there is no feedback between the dynamical variables and the structural properties. This highlights the novelty introduced by this feedback.

The value of $\tau$ depends on the topology. In particular, depending on how $p$ scales with $N$, eq.(\ref{eq_norm}) leads to
\begin{equation}
\tau=\frac{1}{1+pN}\to
\left\{\begin{array}{lll}
1&pN\to 0\\
(1+d)^{-1}&pN=d\\
0&pN\to \infty
\end{array}\right.
\label{eq_tau0}
\end{equation}
These three regimes for the fitness correspond to three different possibilities for the topology. In particular, they are related to the percolation phase transition driven by the parameter $p$ representing the density of the network. For small values of $p$, the graph is split into many small subgraphs or {\em clusters}, whose size is exponentially distributed. As $p$ increases, these clusters become progressively larger, and at the critical percolation threshold $p_c\approx 1/N$  \cite{AB01,siam} the cluster size distribution is scale--invariant. When $p>p_c$ a large giant cluster appears, whose size is of order $O(N)$ and whose relative size tends to $1$ as $p\to 1$.

Now, it is clear that below the percolation threshold (subcritical regime) the minimum--fitness vertex is in most cases isolated, with no neighbours. Thus it is the only updated vertex, and as a result all fintess values, except the newly replaced one, tend towards $1$. This explains why $\tau\to 1$ in this case. By constrast, if $pN=d$ with finite $d>1$ (sparse regime), then a finite number of vertices is updated, and $\tau$ remains finite as $N\to\infty$. This is precisely the case considered in the random--neighbour variant \cite{BSrn}, that we recover correctly.
Finally, if $pN\to \infty$ (dense regime), then an infinite 
number $\langle k_{min}\rangle=pN$ of fitnesses is continuously 
udpated. Therefore $\rho(x)$ is uniform between $0$ and $1$ as in the initial state, and $\tau\to 0$.
Therefore the possible dynamical regimes are tightly related to the topology of the network, depending on the parameter $z$. This result holds also for the more complicated case that we consider below.\\

The average fitness $\langle x\rangle$ can be computed analytically as
\begin{equation}
\langle x\rangle\equiv\int_0^1 x\rho(x)dx
\label{eq_meandef}
\end{equation}
Using eq.(\ref{eq_rhop}) and eq.(\ref{eq_tau0}), it is straightforward to show that
\begin{equation}
\langle x\rangle=\frac{1+\tau}{2}
\label{eq_meanp}
\end{equation}
so that, as expected, $\langle x\rangle$ increases from the initial value $1/2$ to a stationary value linearly dependent on $\tau$.

\subsection{Fitness--dependent networks with local properties}
We now present the analytical solution obtained with the simplest nontrivial choice discussed in section \ref{choice}, $f(x,y)=zxy/(1+zxy)$. 
With the above choice, it can be shown \cite{naturelett} that in the $N\to\infty$ limit eq.(\ref{eq_rho2}) becomes equivalent to
\begin{equation}
\rho(x)=\left\{\begin{array}{ll}
(\tau N)^{-1} &x<\tau\\
(\tau N)^{-1}+2/(zN\tau^2x) &x>\tau
\end{array}\right.
\label{eq_rhofermi}
\end{equation}
Remarkably, now $\rho(x)$ is found to be the 
superposition of a uniform distribution and a power--law with exponent $-1$.
The value of $\tau$, obtained as the solution of eq.(\ref{eq_norm}), reads
\begin{equation}
\tau=\sqrt{\frac{\phi(zN)}{zN}}\to
\left\{\begin{array}{lll}
1&zN\to 0\\
\sqrt{\phi(d)/d}&zN=d\\
0&zN\to \infty
\end{array}\right.
\label{eq_tau}
\end{equation}
As for random graphs, these three dynamical regimes are found to be related to an underlying topological percolation transition \cite{naturelett}.

\begin{figure}
\centerline{
\resizebox{0.9\columnwidth}{!}{%
  \includegraphics{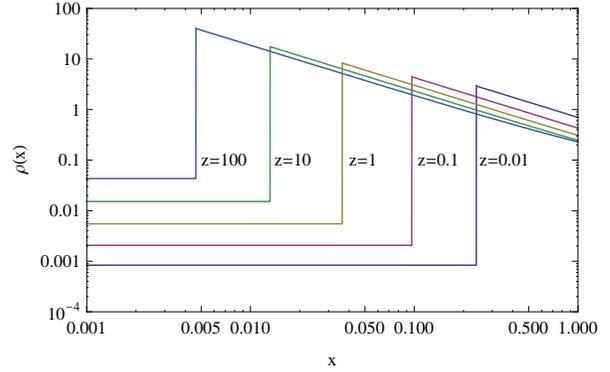}
}
}
\caption{Stationary fitness distribution $\rho(x)$ for $N=5000$ and various values of $z$.}
\label{fig_rho}
\end{figure}
Figure \ref{fig_rho} shows the stationary fitness distribution $\rho(x)$ appearing in eq.(\ref{eq_rhofermi}).  
The theoretical results are in excellent agreement with the numerical simulations shown previously in Fig.\ref{fig:1bis}. 
Also, the theoretical dependence of $\tau$ on $zN$ appearing in eq.(\ref{eq_tau}) is plotted in Fig.\ref{fig_tau} as a solid line, and shown to fit the simulation results perfectly.\\

We can now obtain the exact expression for the average fitness value $\langle x\rangle$ 
at the stationary state. Inserting eq.(\ref{eq_rhofermi}) into eq.(\ref{eq_meandef}) we find 
\begin{equation}
\langle x\rangle=\frac{1}{2\tau N}+\frac{2}{zN\tau^2}(1-\tau)=\frac{1}{2\tau N}+\frac{\tau-1}{\log\tau}
\label{eq_mean}
\end{equation}
where in the last passage we have used the expression $zN=-\log(\tau^2)/\tau^2$ obeyed by $zN$ \cite{naturelett}, representing the inverse of eq.(\ref{eq_tau}). 
For fixed $N$, eq.(\ref{eq_mean}) is plotted in Fig.\ref{fig_mean} as a function of $\tau$. Again the accordance with simulations is very good.
Remarkably, we find that, even if during the evolution $\tau$ increases from $0$ to its asymptotic value as in the case $f(x,y)=p$, now the stationary average value $\langle x\rangle$ is not necessarily larger than the initial value $1/2$. Indeed, for the interesting parameter range, $\langle x\rangle<1/2$. This is the effect of $\rho(x)$ being no longer constant for $x>\tau$.

\begin{figure}[h]
\centerline{
\resizebox{0.9\columnwidth}{!}{%
  \includegraphics{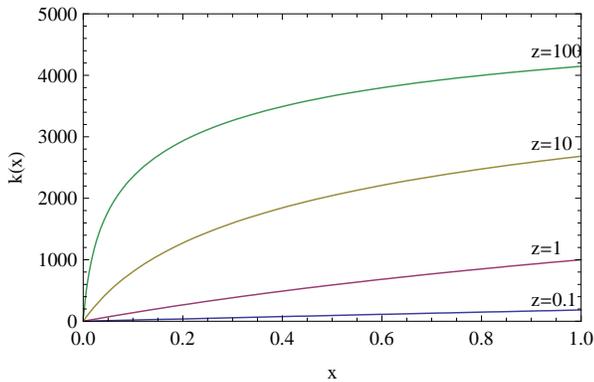}
}
}
\caption{Plot of $k(x)$ for $N=5000$ and various values of $z$.}
\label{fig4}
\end{figure}
In this case, the expected degree $k(x)=N\int f(x,y)\rho(y)dy$ of a vertex with fitness $x$ reads \cite{naturelett}
\begin{equation}
k(x)=\frac{2}{z\tau^2}\ln\frac{1+zx}{1+z\tau x}+\frac{zx-\ln(1+zx)}{z\tau x}
\label{eq_k}
\end{equation}
This behaviour is reported in Fig.\ref{fig4}. For small (but larger than $\tau$) values of $x$, $k$ is proportional to $x$. This implies that in this linar regime the degree distribution $P(k)$ follows the fitness distribution $\rho(x)$, as we reported numerically in section \ref{degree}. By contrast, for large values of $x$ a saturation to a maximum degree is observed. This explains the upper cut--off of the degree distribution. The width of the linear regime decreases as $z$ increases, as the network becomes denser and structural correlations stronger \cite{newman_origin,likelihood,maslov}.
This behaviour can be further characterized analytically \cite{naturelett}, by using the inverse function $x(k)$ to obtain the degree distribution as $P(k)dk= \rho[x(k)]dx$ \cite{fitness}. The analytical form of $P(k)$ perfectly agrees with the numerical results \cite{naturelett}.

\section{Conclusions}
We have discussed in detail a recent model \cite{naturelett} where the interplay between topology and dynamics in complex networks is introduced explicitly. The model is defined by coupling the Bak--Sneppen model of fitness evolution and the fitness model for network formation. The model can be solved analytically for any choice of the connection probability. Remarkably, the fitness distribution $\rho(x)$ self--organizes spontaneously to a stationary probability density, thus removing the need to 
specify an \emph{ad hoc} distribution as in the static fitness model. Moreover, the stationary state is nontrivial and differs from what is observed when dynamics and topology are decoupled. Besides providing a possible explanation for the spontanous emergence of complex topological properties in real networks, these results indicate that adaptive webs offer a new framework wherein unexpected effects can be observed.
                                                                                          
GC acknowledges D. Donato for helpful discussions.
This work was partly supported by the European Integrated Project DELIS.
%
%
%
%
%

\end{document}